# A CF-Based Randomness Measure for Sequences

Anvesh Aileni

**Abstract.** This note examines the question of randomness in a sequence based on the continued fraction (CF) representation of its corresponding representation as a number, or as D sequence. We propose a randomness measure that is directly equal to the number of components of the CF representation. This provides a means of quantifying the randomness of the popular PN sequences as well. A comparison is made of representation as a fraction and as a continued fraction.

## 1. Introduction

Randomness measures are fundamental to many problems in information and cryptography. The question of algorithmic randomness has an old history, going back to Kolgomorov [1]. In this approach a sequence that requires a longer algorithm to generate it is considered to be more random. Several practical approaches to quantifying randomness using transforms of one kind or the other have been proposed in the literature [2],[3].

In this note we consider randomness measures that are especially suited for PN and D sequences, although this applies to any periodic sequence that can be mapped to a D sequence. PN sequences are widely used in cryptography and privacy, simulation, communications, and as random sequences [4],[5]. A linear shift register sequence, with *n*-stages, will have a period of $2^n-1$. PN sequences satisfy many statistical tests of random which is why they remain popular. D sequences do not have as good autocorrelation properties as PN sequences, but they provide great flexibility in terms of the sequence period and this has led to many applications [6]-[11]. Any periodic sequence can be represented as a generalized D sequence *m/n*, where *m* and *n* are suitable natural numbers, i.e., positive integers.

A sequence is said to be more random if it requires a longer program to generate it [1]. For example consider 0000011111 and 1101011001. In these two sequences, the second one looks more random intuitively. If the generator is to be a program in natural language, then the first sequence is simply a string of 0s followed by an equal number of 1s, whereas there is no similar compact generator that one can think of for the second sequence. Thus the intuitive idea of the second sequence being more random than the first one has a logical basis.

In an algorithmic approach to randomness, the size of the generator would be a measure of the randomness. Gangasani [12] used the randomness measure $R(x) = 1 - \dfrac{\sum_{k=1}^{n-1} |C(k)|}{n-1}$, where *C(k)* is the autocorrelation value for *k* and *n* is the period of sequence to characterize the randomness of a sequence. Here, we consider a more intrinsic approach by comparing the generator size for a few PN sequences and their equivalent D sequences. Two kinds of generators are considered: the mathematical representation in terms of the polynomial or rational number for each and that of a continuous fraction (CF) representation. Although the randomness of a PN sequence may be characterized by the degree of its polynomial, here we will argue that a measure based on CF representation is a more general measure that may be applied to any sequence.

2. **PN Sequence Generation**

The PN sequences for polynomials of degree r have been calculated using Matlab's Pseudo Random number generator function **"seqgen.pn"**.

Assume for a polynomial of degree 6, the equation would be $x^6+x+1$ or [6 1 0] or [1 0 0 0 0 1 1]. The PN sequence can be generated using

    h = seqgen.pn('GenPoly', [6 1 0], 'MakeOrShift', [1 1 0 1 0 1])
        or
    h = seqgen.pn('GenPoly', [1 0 0 0 0 1 1], 'Shift', 0)
        or
    h = seqgen.pn('GenPoly', [6 1 0], 'Shift', 0)

then,

    set(h, 'NumBitsOut', $2^6$ -1);
    generate(h)

This generates the PN sequence with a period of $2^6-1$.

The PN sequence may be mapped to a D sequence readily. Assume the PN sequence to be

    1 0 0 1 1 1 0

This is the PN Sequence for polynomial of degree 3. If we write it as the binary sequence 0.1001110, it corresponds to:

    Numerator: $(0 \times 2^0)+(1 \times 2^1)+(1 \times 2^2)+(1 \times 2^3)+(0 \times 2^4)+(0 \times 2^5)+(1 \times 2^6) = 78$
    Denominator: $(2^7) - 1 = 127$

Thus the D Sequence for above PN sequence is equal to $\frac{78}{127}$.

The binary D sequence for $\frac{m}{q}$ is generated by means of the algorithm [3]:

$a(i) = m\ 2^i \bmod q \bmod 2$

where $q$ is a prime number. The maximum length (q-1) sequences are generated when 2 is a primitive root of $q$. When the binary D sequence is of maximum length, then bits in the second half of the period are the complements of those in the first half. The binary sequence $\frac{5}{7}$ is thus

     101

after which the sequence repeats itself. As the numerator changes, for a maximum length D sequence, one obtains a shifted sequence. The period mod 7 is 3 as the order of element 2 modulo 7 is 3.

Table 1 presents a few PN sequences and the corresponding D Sequence written as fractions.

**Table 1:** PN Sequences and corresponding D-Sequences

| PN_SEQ Degree | Polynomial | Equivalent Fraction |
|---|---|---|
| 2 | $x^2+x+1$ | $\frac{5}{7}$ |
| 3 | $x^3+x^2+1$ | $\frac{78}{127}$ |
| 4 | $x^4+x^3+1$ | $\frac{18348}{32767}$ |
| 5 | $x^5+x^3+1$ | $\frac{1119559476}{2147483647}$ |
| 6 | $x^6+x^5+1$ | $\frac{4754309678505905152}{9223372036854775807}$ |

It may be noted that the decimal representations of the PN sequences are the ones obtained by the MATLAB generator and they can be made more efficient by shifting the bits. Each such shift would correspond to division by 2 of the numerator. In other words, $\frac{78}{127}$ and $\frac{39}{127}$ represent the same PN sequence.

### 3. Continued Fraction Representation

Continued fraction (CF) representations go back to Euclid and Indian mathematicians such as Aryabhata [13]-[16]. Let us consider continued fraction representations of the generators for the D sequences corresponding to the given PN sequences. The idea of continued fraction is a natural one to use since such fractions are independent of the radix associated with the number.

The continued fraction for $\frac{78}{127}$, i.e., for $x^6+x+1$, is

$$\cfrac{1}{1+\cfrac{1}{1+\cfrac{1}{1+\cfrac{1}{1+\cfrac{1}{2+\cfrac{1}{4+\cfrac{1}{2}}}}}}}$$

This may be written compactly as [1, 1, 1, 1, 2, 4, 2] and it shows that we need a vector of 7 elements to represent this fraction. The idea is to associate the length of the CF representation to the supposed randomness of the sequence. Formally,

Randomness Measure, $R$ = Number of components of the CF representation

Thus the randomness measure of the sequence $\frac{78}{127}$ is 7. The randomness measure R may similarly be computed for other PN sequences of Table 1.

The randomness measure may be normalized although we will not do so in this paper.

The following table shows the continued fractions for the polynomials of Table 1 and the equivalent fractions.

**Table 2.** Continued Fraction for equivalent fraction of a polynomial

| Polynomial | Equivalent Fraction | Continued Fraction |
|---|---|---|
| $x^2+x+1$ | $\frac{5}{7}$ | [1,2,2] |
| $x^3+x^2+1$ | $\frac{78}{127}$ | [1,1,1,1,2,4,2] |
| $x^4+x^3+1$ | $\frac{18348}{32767}$ | [1,1,3,1,2,34,7,1,2] |
| $x^5+x^3+1$ | $\frac{1119559476}{2147483647}$ | [1,11,4,1,1,2,11,12,6,12,1,16,2,2,3] |
| $x^6+x^5+1$ | $\frac{4754309678505905152}{9223372036854775807}$ | [1,1,15,1,3,174,1,15,17,1,1,1,1,23,1,1,5, 1,4,34,1,1,1,1,1,2,1,18,3,1,1,7,1,1,84] |

The size of the numerator and the size of the CF vector are roughly proportional although the number of components of the CF vector is higher.

**More on Continued fractions:**

A few observations were made on the fractions with 127 as the denominator. For these fractions, the length of the continued fraction is calculated and binary sequence for the fraction was computed.

In Table 3 we consider the fraction together with corresponding continued fraction, length and binary sequence. It may be observed from the table that the binary sequence with less random sequence has smaller length for continued fraction.

For example, the length of continued fraction for a binary sequence 0000111 corresponding to $\frac{7}{127}$ is 2, similarly for 0011111 corresponding to $\frac{31}{127}$ has 3, and for 011111 $\frac{63}{127}$ has 2, while for 1101001 corresponding to $\frac{105}{127}$ which is more random has 6. This seems to agree with the intuitive notion that a sequence with alternating subsequences of 0s and 1s with no specific size are more random than those where all the 0s and 1s are clumped separately.

**Table 3:** Table showing length if continued fraction and binary sequence for fractions

| S. No | Fraction | Continued Fraction | Length=$R$ | Binary Sequence |
|---|---|---|---|---|
| 1 | 3/127 | [42, 3] | 2 | 0000011 |
| 2 | 7/127 | [18, 7] | 2 | 0000111 |
| 3 | 13/127 | [9, 1, 3, 3] | 4 | 0001101 |
| 4 | 15/127 | [8, 2, 7] | 3 | 0001111 |
| 5 | 19/127 | [6, 1, 2, 6] | 4 | 0010011 |
| 6 | 20/127 | [6, 2, 1, 6] | 4 | 0010100 |
| 7 | 25/127 | [5, 12, 2] | 3 | 0011001 |
| 8 | 31/127 | [4, 10, 3] | 3 | 0011111 |
| 9 | 33/127 | [3, 1, 5, 1, 1, 2] | 6 | 0100001 |
| 10 | 39/127 | [3, 3, 1, 9] | 4 | 0100111 |
| 11 | 45/127 | [2, 1, 4, 1, 1, 1, 2] | 7 | 0101101 |
| 12 | 47/127 | [2, 1, 2, 2, 1, 4] | 6 | 0101111 |
| 13 | 57/127 | [2, 4, 2, 1, 1, 2] | 6 | 0111001 |
| 14 | 63/127 | [2, 63] | 2 | 0111111 |
| 15 | 77/127 | [1, 1, 1, 1, 5, 1, 3] | 7 | 1001101 |
| 16 | 78/127 | [1, 1, 1, 1, 2, 4, 2] | 7 | 1001110 |
| 17 | 79/127 | [1, 1, 1, 1, 1, 4, 1, 2] | 8 | 1001111 |
| 18 | 81/127 | [1, 1, 1, 3, 5, 2] | 6 | 1010001 |
| 19 | 97/127 | [1, 3, 4, 3, 2] | 5 | 1100001 |
| 20 | 105/127 | [1, 4, 1, 3, 2, 2] | 6 | 1101001 |
| 21 | 107/127 | [1, 5, 2, 1, 6] | 5 | 1101011 |

The 21 sequences shown in Table 3 are mapped into Graph 1 that shows how the randomness measure varies across the set of sequences. The randomness value is small only for the sequences where the 0s and 1s are separately clumped.

**Graph 1:** Graph representing length of continued fraction

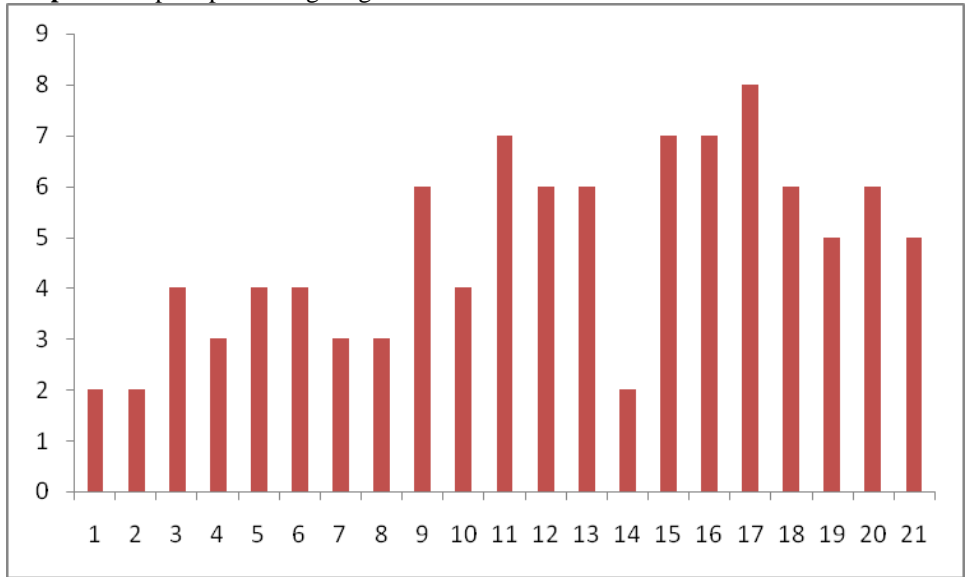

In Tables 4 and 5 we contrast the CF randomness measure for "structured" and "unstructured" sequences. In Table 4, the binary sequences have clear periodic structure and the CF length is either 1 or 2 as one would expect intuitively. Conversely, in Table 5, the randomness measure varies between 8 and 13, which is once again in conformity with our intuitive expectation.

**Table 4:** Table showing length of continued fraction for less random binary sequence

| SNo | Binary Sequence  | Fraction    | Continued Fraction | Length=$R$ |
|-----|------------------|-------------|--------------------|------------|
| 1   | 1111000011110000 | 61680/65535 | [1, 61680]         | 2          |
| 2   | 0000111100001111 | 3855/65535  | [3855]             | 1          |
| 3   | 1111111100000000 | 65280/65535 | [1, 65280]         | 2          |
| 4   | 0000000011111111 | 255/65535   | [255]              | 1          |
| 5   | 1100110011001100 | 52425/65535 | [1, 52425]         | 2          |
| 6   | 0011001100110011 | 13107/65535 | [13107]            | 1          |
| 7   | 1010101010101010 | 43690/65535 | [1, 43690]         | 2          |
| 8   | 0101010101010101 | 21845/65535 | [21845]            | 1          |

Furthermore, it can be observed from the table that the length of continued fraction vary based on the randomness in the sequence of the 16 bit binary digits.

One interesting point which may be observed from the table is that the complements have difference of 1 in the length of the continued fraction. For example, 1000111101011000 has R=10 as the length of continued fraction and for 0110101101010010 which is the complement for earlier mentioned sequence has R=9 as its length. Also, the binary sequence when represented in decimal format, the larger number has continued fraction 1 more than the lesser number.

**Table 5:** Table showing length of continued fraction for random 16 bit binary sequence

| SNo | Binary Sequence | Fraction | Continued Fraction | Length = $R$ |
|---|---|---|---|---|
| 1 | 1000111101011000 | 36696/65535 | [1, 1, 3, 1, 2, 28, 1, 3, 3, 2] | 10 |
| 2 | 0110101101010010 | 28839/65535 | [2, 3, 1, 2, 28, 1, 3, 3, 2] | 9 |
| 3 | 1001110100011110 | 44402/65535 | [1, 2, 9, 1, 8, 2, 2, 3, 1, 2, 1, 2] | 12 |
| 4 | 0111000010100111 | 21133/65535 | [3, 9, 1, 8, 2, 2, 3, 1, 2, 1, 2] | 11 |
| 5 | 1001110100011110 | 40222/65535 | [1, 1, 1, 1, 2, 3, 4, 3, 6] | 9 |
| 6 | 0101001010001101 | 25313/65535 | [2, 1, 1, 2, 3, 4, 3, 6] | 8 |
| 7 | 1010100010101101 | 43181/65535 | [1, 1 ,1, 13, 1, 1, 1, 3, 2, 1, 2, 7, 2] | 13 |
| 8 | 0101011101010010 | 22354/65535 | [2, 1, 13, 1, 1, 1, 3, 2, 1, 2, 7, 2] | 12 |
| 9 | 1000101001100101 | 35429/65535 | [1, 1, 5, 1, 1, 1, 9, 1, 1, 2, 3, 2, 4] | 13 |
| 10 | 0111010110011010 | 30106/65535 | [2, 5, 1, 1, 1, 9, 1, 1, 2, 3, 2 ,4] | 12 |
| 11 | 1001010010101101 | 38061/65535 | [1, 1, 2, 1, 1, 2, 7, 1, 2, 2, 12] | 11 |
| 12 | 0110101101010010 | 27474/65535 | [2, 2, 1, 1, 2, 7, 1, 2, 2, 12] | 10 |

## 4. Conclusions

This article has proposed the use of continued fractions to represent the randomness of a sequence. It is suggested that a sequence that has a longer continued fraction representation is to be considered more random. This provides a suitable means of quantifying the randomness of D sequences and PN sequences. A comparison is made of representation as a fraction and as a continued fraction for several cases.

Experiments have been done with several PN sequences and with other random sequences. The proposed measure appears superior to that which considers only the size of the numerator in the D sequence.